\documentstyle[epsfig,color]{mn}

\def\ltsim{\mathrel{\hbox{\rlap{\hbox{\lower4pt\hbox{$\sim$}}}\hbox{$<$}}}}
\def\gtsim{\mathrel{\hbox{\rlap{\hbox{\lower4pt\hbox{$\sim$}}}\hbox{$>$}}}}

\title{Young star clusters in the Large Magellanic Cloud: NGC 1805 and NGC 1818}

\author[Johnson et al.]
{R.A.Johnson$^1$,
S.F.Beaulieu$^1$,
G.F.Gilmore$^1$,
J.Hurley$^1$,
B.X.Santiago$^2$,\cr
N.R.Tanvir$^3$, 
R.A.W.Elson$^4$
\\
$^1$Institute of Astronomy, Madingley Road, Cambridge CB3 0HA,
United Kingdom \\ email: raj@ast.cam.ac.uk \\
$^2$Instituto de Fisica, Universidade Federal Rio Grande do Sul, 
91510-970 Porto Alegre, RS Brasil\\
$^3$Department of Physical Science, University of Hertfordshire, College
Lane, Hatfield, AL10 9AB\\
$^4$Deceased}

\begin{document}

\maketitle

\begin{abstract}

We present colour-magnitude diagrams for two rich 
($\approx$10$^{4}$M$_{\sun}$) Large Magellanic Cloud star clusters with ages
$\approx$10$^7$ years, constructed from optical and near-infrared data
obtained with the Hubble Space Telescope\footnotemark. These data are
part of an HST project to study LMC clusters with a range of ages.  In
this paper we investigate the massive star content of the young
clusters, and determine the cluster ages and metallicities, paying
particular attention to Be star and blue straggler populations and
evidence of age spreads. We compare our data to detailed stellar
population simulations to investigate the turn-off structure of
$\approx$25 Myr stellar systems, highlighting the complexity of the
blue straggler phenomenon.

\end{abstract}


\footnotetext{"Based on observations with the NASA/ESA Hubble Space
Telescope, obtained at the Space Telescope Science Institute, which is
operated by the Association of Universities for Research in Astronomy,
Inc. under NASA contract No. NAS5-26555."}

\section{Introduction}
\label{sec:intro}

Unlike the Milky Way, the Magellanic Clouds contain a population of
young (10$^{6}$-10$^{7}$ yr) rich star clusters, sometimes called
populous clusters. These clusters have been popular targets for 
studies of stellar evolution and cluster formation. With HST, deep
colour-magnitude diagrams can be produced, even of the dense central
regions of the clusters.

Of particular interest is whether these clusters show an age spread
amongst the massive stars, providing evidence for the currently
unknown duration of cluster formation. An age spread amongst the
turn-off stars would appear as a spread in colour larger than the
measured errors.  The age spread in some clusters, e.g. Trapezium
\cite{Pros94}, NGC6531 \cite{Forb96}, is small, only
a few dynamical times. It has been argued \cite{Elme00} that the
star formation timescale is of this order on all spatial scales.  On
the other hand some clusters show evidence for larger age spreads, 
e.g. NGC1805 \cite{Calo98}, and longer star formation timescales are
required in some models of the initial mass function based on clump or
protostar interactions.  The timescale for cluster formation has
implications for the efficiency of star formation. If the star
formation timescale is short ($\sim$ the dynamical time) then, for
the cluster to remain bound, the efficiency must be high. If the star
formation timescale is longer then the efficiency can be lower, as gas
can be removed without disrupting the cluster \cite{Elso87,Geye00}.

The search for age spreads is complicated by the presence in the
clusters of Be stars and blue stragglers.  The Be and blue straggler
populations are also interesting to study in their own right,
especially because the factors affecting the numbers of blue
stragglers and Be stars are not well understood.  The unique young,
metal-poor and rich environment of the young Magellanic clusters
therefore provides a useful addition to the parameter space studied.

Blue stragglers are stars that are bluer and brighter than the main
sequence turn-off. They are thought to be produced by the merger of
two normal cluster stars, either in a primordial binary system, or
through direct stellar collisions, or both.  All Galactic globular
clusters that have been surveyed contain some blue stragglers, though
the fraction of blue stragglers varies greatly, even between similar
clusters.  Some Galactic open clusters also contain blue
stragglers. Blue stragglers are found in open clusters of all ages
\cite{Ahum95}. In the Magellanic Clouds, blue straggler stars have
been found in old (NGC121 \cite{Shar98}, NGC1466 \& NGC2257
\cite{John99}) and young (NGC330 \cite{Kell00}) clusters.

Be stars are non-supergiant (luminosity class V to III) B type stars
that show or once have shown Balmer emission \cite{Jasc81}. It is
widely accepted that the Be phenomenon is associated with rapid
rotation of the stellar photosphere, and the presence of a
circumstellar disk that gives rise to the line emission.  It has also
been observed that Be stars are redder in V-I than normal B stars
\cite{Greb97,Kell99}, and that those with the strongest H$\alpha$
emission are also the reddest e.g. \cite{Dach88}.
1988). The reason for this reddening is not entirely clear, but it
seems likely that continuum emission from the disk plays a large part.
There is also possibly some effect from a change in spectral energy
distribution due to rotational distortion of the stellar atmosphere.
Keller et al.\ \shortcite{Kell00} find that the Be star fraction peaks at
the main sequence turn-off, which suggests that the Be star phenomenon
occurs at a specific evolutionary stage. There are however
differences in Be star fraction amongst clusters of the same age, and
some evidence that metallicity also affects the Be star fraction
\cite{Maed99}.

The two young clusters discussed in this paper are the LMC clusters,
NGC 1805 and NGC 1818. Both clusters are located in the north west
part of the LMC, $\sim 3.2$ kpc from the centre, in fairly low density
regions.  Previous measurements of the cluster parameters are
summarized in Table~\ref{tab:clusprops}.

\begin{table}
\caption{Properties of LMC clusters NGC1805 and NGC1818}
\begin{tabular}{llll} \hline
\multicolumn{4}{l}{NGC1818} \\ \hline
Position (J2000)   & \multicolumn{2}{l}{5:04:14, $-$66:26:05} & \\
M$_V$              & -8.8 & & 1 \\
Mass (M$_{\sun}$)   & $3 \times 10^4$ & & 2 \\
Metallicity [Fe/H] & $\approx$-0.4$^{*}$ & spectroscopy & 3,4,5 \\
Age (Myr)          & 20-40    & ground-based images & 6 \\
                   & 20       & HST images          & 2 \\
Reddening E(B-V)   & 0.1 & spectroscopy & 4 \\
                   & 0.07 & IUE spectroscopy & 7 \\
                   & 0.05 & integrated IUE colours & 8 \\
                   & 0.07 & ground based images & 6 \\
                   & 0.05 & HST images & 2 \\
log(Dyn. time) (Myr) & 6.2-7 & & 9 \\ \hline
\multicolumn{4}{l}{NGC1805} \\ \hline
Position (J2000)   &\multicolumn{2}{l}{5:02:21, $-$66:06:44} & \\
M$_V$              & -7.9 & & 1 \\ 
Mass ($M_\odot$)   & $6 \times 10^3$ & &  \\ \hline
\end{tabular}
1 van den Bergh \shortcite{Berg81} \hspace*{2cm} * see text\\
2 Hunter et al.\ \shortcite{Hunt97} \\
3 J\"{u}ttner \shortcite{Jutt93a} \\
4 J\"{u}ttner et al.\ \shortcite{Jutt93b} \\
5 Korn et al.\ \shortcite{Korn00}\\
6 Will et al.\ \shortcite{Will95} \\
7 Cassatella et al.\ \shortcite{Cass87} \\
8 Meurer et al.\ \shortcite{Meur90} \\
9 Elson et al.\ \shortcite{Elso87}
\label{tab:clusprops}
\end{table}
Prior to this work, no colour-magnitude diagram (CMD), ground based or
other, existed for NGC 1805.  NGC 1818 has been relatively well
studied.  There have been several attempts to measure the abundances
of stars in NGC1818.  High resolution spectroscopy of LMC cluster and
field stars \cite{Korn00,Jutt93a,Jutt93b} has found similar abundances
for the clusters and the field.  It is likely therefore that NGC1818
and NGC1805 have metallicities of [Fe/H]$\approx$-0.4 (the canonical
value for the LMC). However, we note that no abundance measurements
exist for NGC1805, and that there are only two stars with reliably
measured abundances in NGC1818 (stars D1 and D12 in Korn et al.\ 2000
and D12 in J\"{u}ttner et al.\ 1993).  Previous HST observations of
NGC1818 have been used to derive a mass function \cite{Hunt97}
investigate the binary fraction and mass segregation \cite{Elso98},
and search for Be stars and blue stragglers \cite{Kell00}.

The new observations in this paper are part of the large HST project
GO7307, details of which can be found in Beaulieu et al.\
\shortcite{Beau00}.

The format of this paper is as follows: Section~\ref{sec:data}
describes the data and reductions, Section~\ref{sec:res} presents the
results and these are compared with simulations in Section~\ref{sec:sim}.

\section{Observations and Data Reduction}
\label{sec:data}
\begin{figure*}

{\large top) Available as 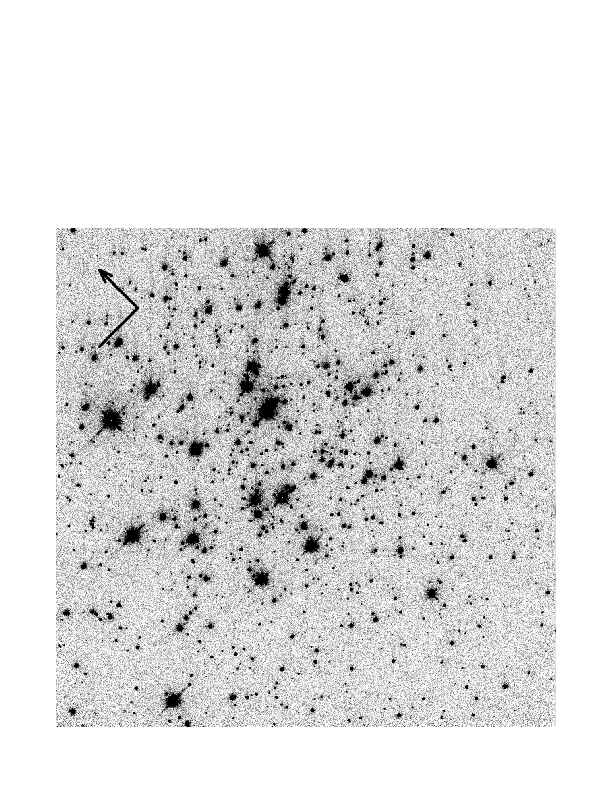}
\vspace*{110mm}

{\large bottom) Available as 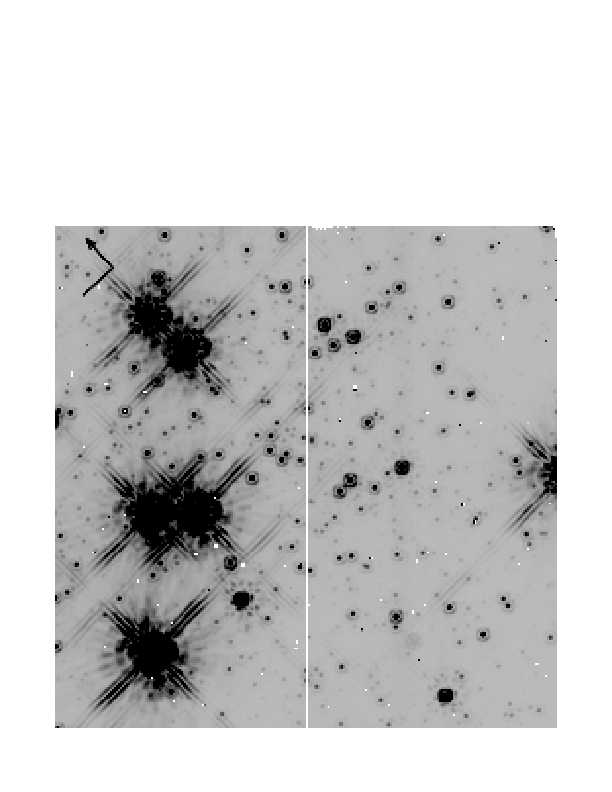}
\vspace*{110mm}
\caption{Images of the core of NGC 1818.
top) PC from the short exposure (15s) in F555W bottom) NICMOS2 with the F160W
passband.  The PC field size is $33 \times 33$ arcsec and the NICMOS2
field size is $19 \times 19$ arcsec.}
\label{fig:n1818pics}
\end{figure*}

\begin{figure*}

{\large top) Available as 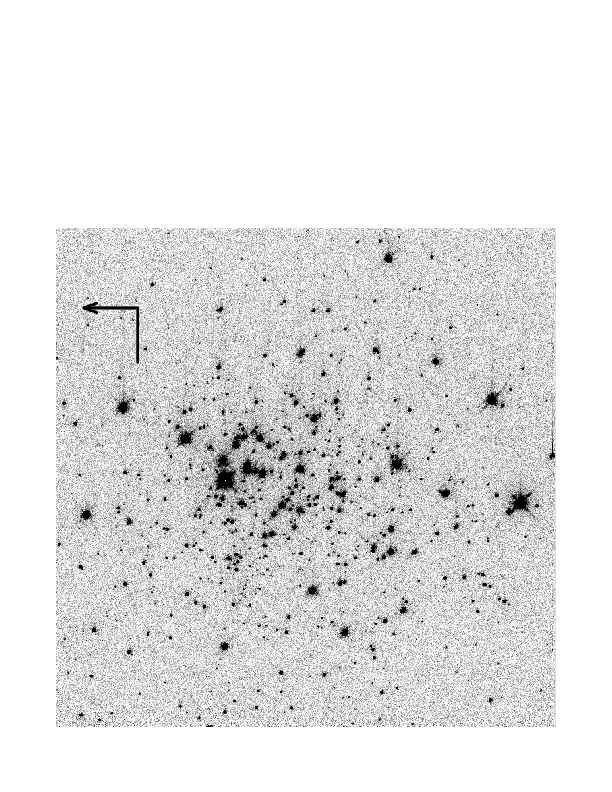}
\vspace*{110mm}

{\large bottom) Available as 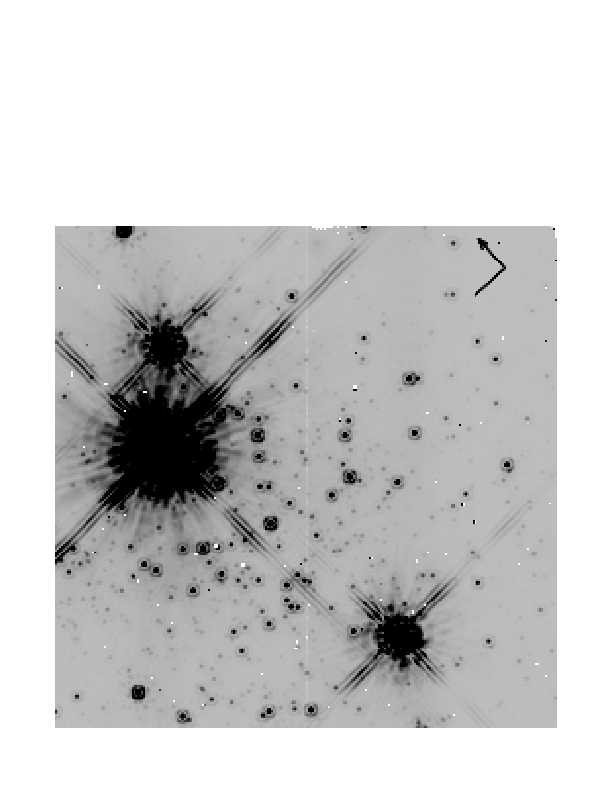}
\vspace*{110mm}
\caption{Images of the core of NGC 1805.
top) PC from the short (15s) exposure in F555W bottom) NICMOS2 with the F160W
passband.  The PC field size is $33 \times 33$ arcsec and the NICMOS2
field size is $19 \times 19$ arcsec.}
\label{fig:n1805pics}
\end{figure*}

For both clusters we obtained HST WFPC2 F555W ($\approx$V) and F814W ($\approx$I) and NICMOS Camera 2 F160W ($\approx$H) observations.
Our original intention was to use the larger field of view NICMOS
Camera 3, but unfortunately this was out of focus during NICMOS' lifetime.
The images discussed here were centred on the clusters, and the
PC and NIC2 contain most of the cluster cores (diameters $\ltsim
20$ arcsec).  We obtained both short exposures to avoid saturating the
brightest stars, and long exposures to provide good signal-to-noise
well below the main-sequence turnoffs.  Full details of the image set
are given in Table~\ref{tab:imdetails}. 
\begin{table}
\caption{The image set for NGC1818 and NGC1805}
\begin{tabular}{llll} \hline
\multicolumn{4}{l}{WFPC2} \\
filter&\multicolumn{2}{c}{dataset} & time (s)\\ \hline
      & NGC1818   & NGC1805   & \\
F555W & u4ax3001r & u4ax0201r & 5 \\
      & u4ax3002r & u4ax0202r & 5 \\
      & u4ax3003r & u4ax0203r & 5 \\
      & u4ax3004r & u4ax0204r &140 \\
      & u4ax3005r & u4ax0205r &140 \\
      & u4ax3006r & u4ax0206r &140 \\	
F814W & u4ax3007r & u4ax0207r &20 \\
      & u4ax3008r & u4ax0208r &20 \\
      & u4ax3009r & u4ax0209r &20 \\
      & u4ax300ar & u4ax020am &300 \\
      & u4ax300bm & u4ax020br &300 \\
      & u4ax300cr & u4ax020cr &300 \\ \hline
\multicolumn{4}{l}{NICMOS2} \\
filter&\multicolumn{2}{c}{dataset} & time (s)\\ \hline
      & NGC1818   & NGC1805   & \\
F160W & n4ax29oaq & n4ax01ahq & 160 \\
      & n4ax29okq & n4ax01arq & 160 \\
      & n4ax29omq & n4ax01atq & 514 \\
      & n4ax29ooq & n4ax01avq & 514 \\
      & n4ax29orq & n4ax01ayq & 514 \\
      & n4ax29otq & n4ax01b0q & 514 \\
      & n4ax29owq & n4ax01b3q & 514 \\ \hline
\end{tabular}
\label{tab:imdetails}
\end{table}
Images of the cluster cores
taken with both the PC and NIC2 are shown in Figures~\ref{fig:n1818pics} and \ref{fig:n1805pics}.

\subsection{WFPC2 reductions}
\label{subsec:WFPC2red}

The individual exposures were combined using the IRAF task crrej,
which sums the images and rejects cosmicrays. After combining we
checked the magnitudes of bright stars in the combined and original
images and adjusted the scalenoise parameter if necessary.  Warm
pixels were flagged using the warm pixel lists produced by STScI.

The NGC1805 long data contains earth light which has been reflected
off the Optical Telescope Assembly baffles and secondary support and
into the WFPC2. This reflected light raises the overall background of
the field and often also produces cross shaped or diagonal stripes in
the level of the background where the WFPC2 camera mirror supports
vignette the scattered light. The vignetting is most prominent on the
F814W images, with the worst affected observation being u4ax020am
where the background is reduced by $\approx$30\% in the stripe.

The striping in the F555W and F814W long images was removed from each
individual image before combining by producing an image which just
contained the stripe pattern and subtracting this `pattern image' from
the original image. The background sky level varied from image to
image and so this was subtracted off each image before combining.  The
sum of the backgrounds was added back to the combined image.

Stars were detected on the F814W image using daofind, and their
magnitudes were measured in the F555W and F814W images using both
aperture (of 2 pixel radius) and point spread function (psf) fitting
photometry, using the IRAF tasks phot and allstar
respectively. TinyTim psfs \cite{Kris97} were used for the psf
fitting. The chi and sharpness values produced from the psf fitting,
which measure the goodness of fit of the psf and the difference
between the square of the width of the object and the square of the
width of the psf, were used to eliminate extended objects and spurious
detections such as residual cosmic rays and warm pixels.  The
magnitudes used in the final colour magnitude diagrams are aperture
magnitudes, as it was found that these produced main sequences that
were at least as narrow as those using the psf fitted magnitudes.

Spurious detections are found around bright stars and also along
their diffraction spikes. We therefore removed all detections within
20-30 pixels of the centres of the bright stars and along the
diffraction spikes. This masking was carried out on the F814W image
where the stars are brightest and affect the largest surrounding area.

We have corrected the data for geometric distortion effects
\cite{Holt95a} and charge transfer efficiency (CTE) effects
\cite{Whit99}.  For the short exposure (and hence low background)
images used here the CTE correction can be quite large, especially for
faint stars at high row numbers (eg $\approx$0.08 mags for a star with
V$\approx$16 at 400,700 in the F555W PC short image).

After correcting for CTE effects the magnitudes of stars on the short
and long exposures were compared.  It was found that the brightest
stars common to both images (V$_{555}\approx$17.5) were brighter in
the short image than in the long image.  Since CTE and other related
effects are known to be more of a problem in low background exposures,
it was decided to correct the short data to the long data. The short
data are only used to get the magnitudes of the brightest stars and so
a shift has been added to the short exposure magnitudes so that the
long and short magnitudes of the bright stars are equal. Note that
this discrepancy between long and short exposures is in the opposite
sense to that found by Casertano \shortcite{Case98}.
Table~\ref{tab:shifts} gives the shifts that were added to the short
data.
\begin{table}
\caption{Shifts added to short data to equate long and short magnitudes for bright stars in common.}
\begin{tabular}{lllll} \hline
   &\multicolumn{2}{c}{NGC1818} & \multicolumn{2}{c}{NGC1805} \\
   & F555W & F814W & F555W & F814W \\ \hline
 PC& 0.06  & 0.05  & 0.05  & 0.03  \\
WF2& 0.03  & 0.01  & 0.02  & 0     \\
WF3& 0.04  & 0.02  & 0.01  & 0.01  \\ 
WF4& 0.05  & 0.03  & 0.03  & 0.01  \\ \hline
\end{tabular}
\label{tab:shifts}
\end{table}

To calibrate our magnitudes in the HST instrumental system we
calculated an aperture correction to 0.5 arcsec, varying with radial
distance from the centre of each chip, from bright stars in the image
and used the zeropoints in Baggett \shortcite{Bagg97}.  We have also
transformed the data to the Johnson-Cousins system using the
transformation equations in Holtzman et al.\ \shortcite{Holt95b}. The
data are de-reddened before transforming. The error in the
transformations is $\approx$2\% for stars with V-I$>\approx$0. For
blue stars with V-I$<$0 the scatter in the F814W to I transformation
increases to $\approx$4\%. This means that for V-I$<$0 the systematic
error in the colour due to the transformations could be as much as
0.04 mags.

The final sample for each chip was formed by using the magnitude from
the long exposure image for all unsaturated stars and from the short
image for those stars that are saturated on the long image. The change
from long to short data is at V$\approx$17.5.

In NGC1805 we find that there is a colour shift between the chips of
$\approx$0.04 mag. Similar shifts have been found by other groups in
cluster colour magnitude diagrams \cite{John99} and attributed to
errors in CTE and aperture corrections and in zeropoints. Similar
errors are likely causing the colour shift seen in our data.

\subsection{NICMOS2 reductions}
\label{subsec:nicred}

The NICMOS data were combined using the IRAF task mscombine, 
which sums the data and performs cosmic ray rejection.
Stars were detected in the NICMOS image and aperture photometry
was performed.

Detections near to bright stars and along diffraction spikes
were masked, as in the optical data.

There are several difficulties with NICMOS data, fortunately none
of these had a big effect on this project. The pedestal, a constant
which remains after running the calibration pipeline and causes
an inverse flatfield pattern to be imprinted on the image, is not
a big problem for these data as we take a local background for
each star. Ghosts, which appear at congruent positions in 
the other quadrants when a bright star is present in one quadrant,
do appear in our images, but it is possible to look carefully
at the positions where ghosts are expected to occur and eliminate
false detections.

An aperture correction to 0.5 arcsec was calculated from bright stars in
the image. The data were calibrated to magnitudes in an approximate
Vega system using
\[
m=\it{ZP}_{\rm{Vega}} - 2.5log(\it{PHOTFNU} * \it{Count Rate} * F_{\nu \rm{Vega}}^{-1})
\]
where {\it PHOTFNU} and {\it F}$_{\nu \rm{Vega}}$ are calculated by STScI at the
time of writing to be 2.337E-6 Jy$\times$s/DN and 1039.3 Jy respectively
(NICMOS Data Handbook, v4, Table 5.1), and we assume
$\it{ZP}_{\rm{Vega}}$=0 (as in the CIT infrared photometry scale).

The final detected star lists in NICMOS F160W and WFPC2 F555W were
matched using the positional information in the image headers. It was
found that there can be as much as 2$^{\prime\prime}$ offset between
the RAs and Decs calculated for a star from the NICMOS image and those
calculated for the same star from the WFPC2 image.  According to STScI
this is due to the combined uncertainty of the guide star positions,
the location of the fine guidance sensors relative to the telescope
axis and the measured locations of the instrument apertures.

\section{Results}
\label{sec:res}

\subsection{Colour magnitude diagrams}
\begin{figure*}
\psfig{file=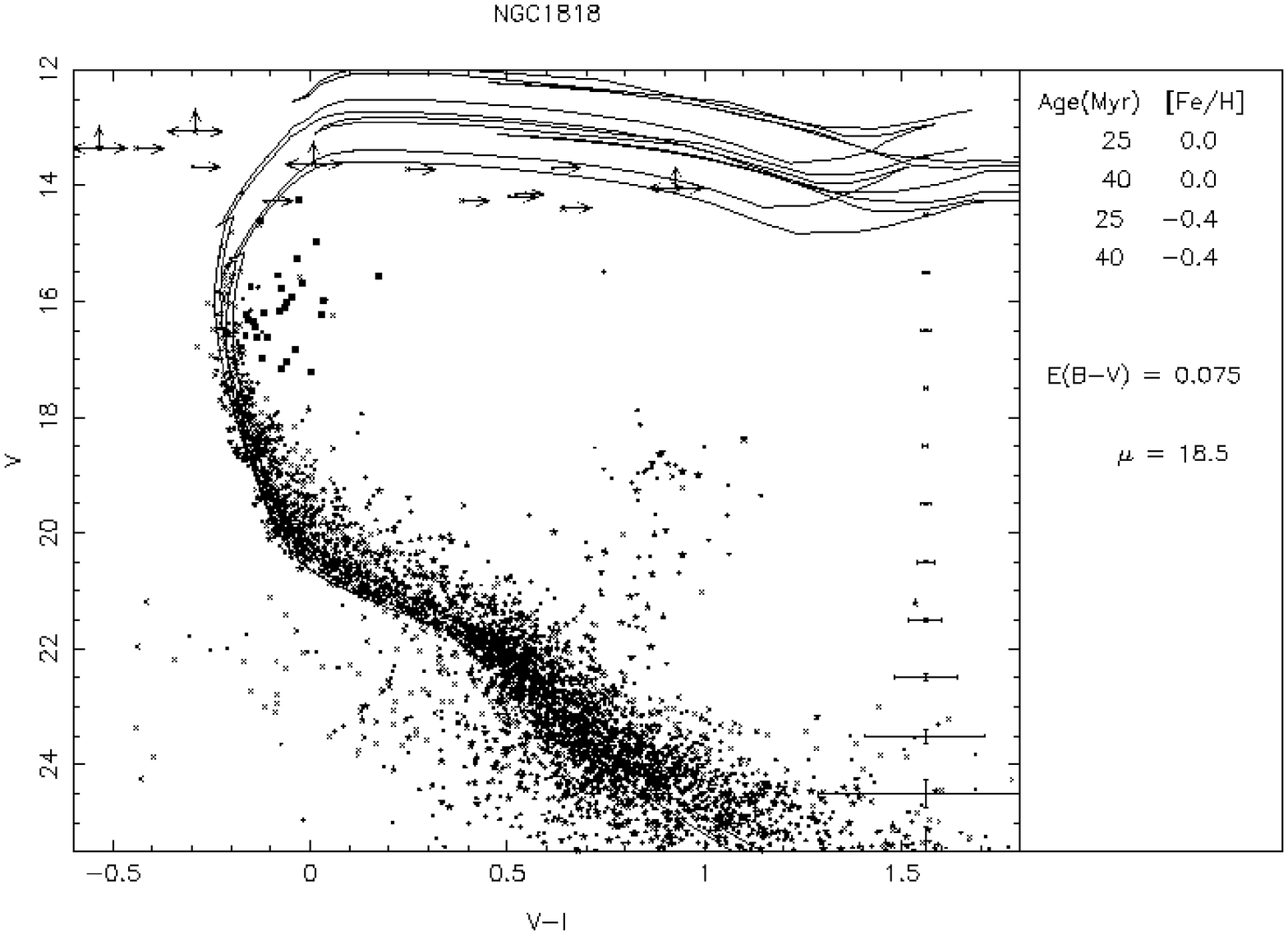,width=11.2cm,angle=-90}
\psfig{file=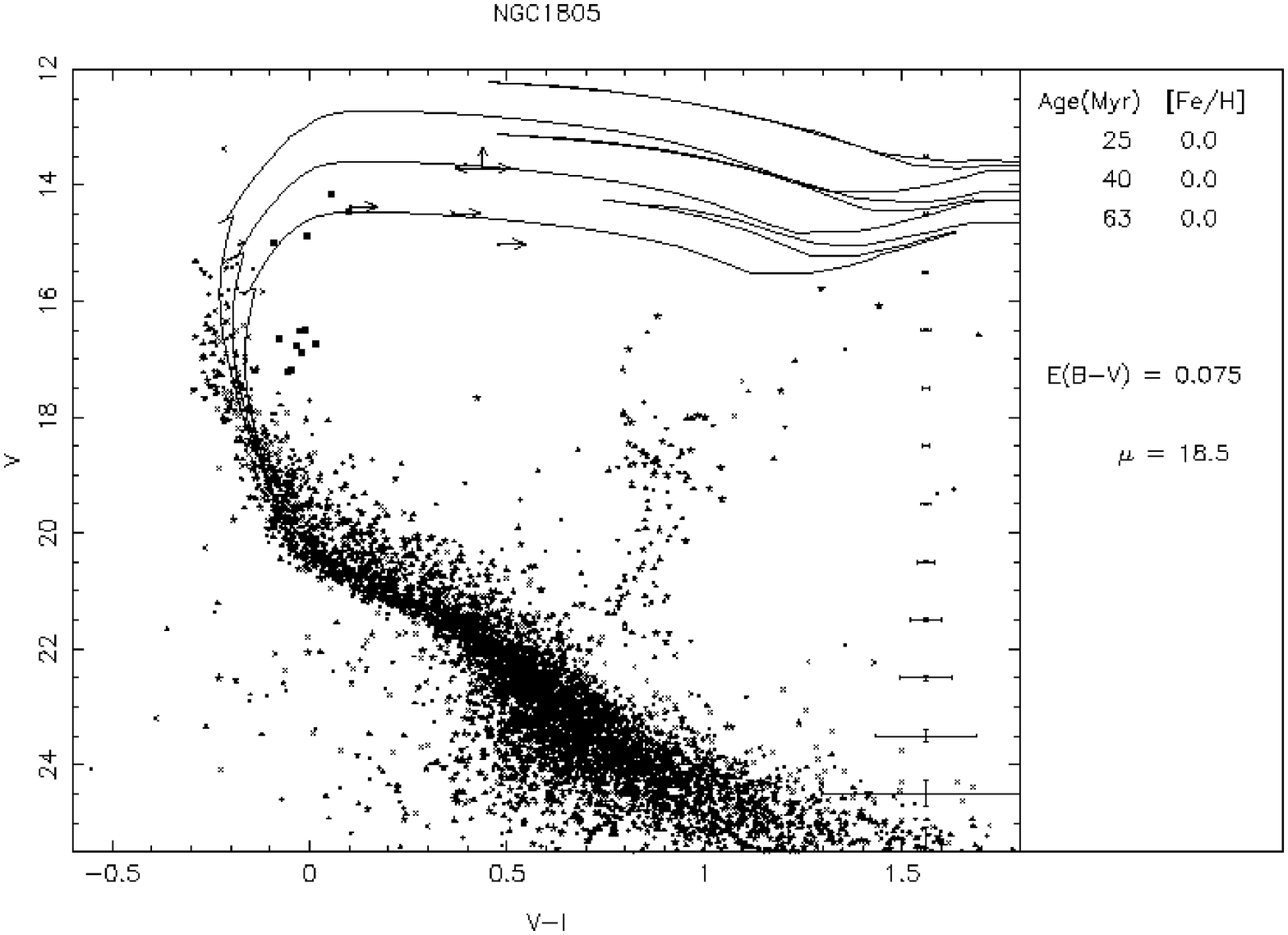,width=11.2cm,angle=-90}
\caption{V vs V-I colour magnitude diagrams for NGC1818 (top) and
NGC1805 (bottom). The data are from all four WFPC2 chips, which are
denoted by the following symbols: PC - cross, WF2 - circle, WF3 -
star, WF4 - triangle. Bold squares mark Be stars. The isochrones shown
are for ages 25 and 40 Myr and metallicities [Fe/H]=-0.4 and 0.}
\label{fig:VIcolmag}
\end{figure*}

Figure~\ref{fig:VIcolmag} shows the de-reddened V vs V-I
(Johnson-Cousins magnitudes) colour-magnitude diagrams (CMDs) for all
four chips of NGC1818 (top) and NGC1805 (bottom).  The different chips
have different symbols. Stars marked with bold squares are Be stars
(see subsection~\ref{subsec:Beid}).
The data have been de-reddened assuming E(B-V)=0.075.
Tables~\ref{tab:n1818dat} and \ref{tab:n1805dat} tabulate these data
for NGC1818 and NGC1805 respectively (full versions of these tables
are available on the MNRAS web site).
\begin{table*}
\caption{NGC1818 WFPC2 data table (a full version is available from the MNRAS web site). The columns contain: 1) Chip no. 1=PC, 2=WF2, 3=WF3, 4=WF4 2) \& 3) x\&y coordinates measured on the F555W image 4) \& 5) RA and Dec 6)-9) Johnson-Cousins V \& I magnitudes and errors obtained
as described in subsection~\ref{subsec:WFPC2red} and de-reddened using E(B-V)=0.075 10) Be star flag (see subsection~\ref{subsec:Beid})}
\label{tab:n1818dat}
\begin{minipage}{18cm}
\begin{tabular}{cccccccccc} \hline 
Chip & x & y &\multicolumn{2}{c}{RA (J2000) Dec} & V & $\Delta$V & I & $\Delta$I & Flag\\ \hline 
1 & 74.31 & 112.20 & 5:04:17.406 & -66:26:04.38 & 23.416 & 0.118 & 22.748 & 0.088 &  \\ 
1 & 110.96 & 118.00 & 5:04:17.183 & -66:26:05.37 & 24.410 & 0.223 & 23.648 & 0.176 &  \\ 
1 & 134.37 & 121.83 & 5:04:17.039 & -66:26:06.00 & 23.321 & 0.096 & 22.797 & 0.116 &  \\ 
1 & 87.85 & 123.43 & 5:04:17.276 & -66:26:04.46 & 20.992 & 0.016 & 20.840 & 0.018 &  \\ 
1 & 98.06 & 124.98 & 5:04:17.214 & -66:26:04.74 & 21.209 & 0.018 & 20.944 & 0.019 &  \\ 
1 & 157.44 & 129.87 & 5:04:16.874 & -66:26:06.49 & 23.172 & 0.102 & 22.429 & 0.098 &  \\ 
1 & 139.30 & 132.88 & 5:04:16.954 & -66:26:05.81 & 21.564 & 0.025 & 21.237 & 0.032 &  \\ 
1 & 185.89 & 133.40 & 5:04:16.705 & -66:26:07.29 & 23.700 & 0.18 & 22.833 & 0.152 &  \\ \hline
\end{tabular}
\end{minipage}
\end{table*}

\begin{table*}
\caption{NGC1805 WFPC2 data table. Columns as in Table~\ref{tab:n1818dat}.}
\label{tab:n1805dat}
\begin{minipage}{18cm}
\begin{tabular}{cccccccccc} \hline 
Chip & x & y &\multicolumn{2}{c}{RA (J2000) Dec} & V & $\Delta$V & I & $\Delta$I & Flag\\ \hline 
1 & 499.83 & 64.34 & 5:02:24.153 & -66:06:47.66 & 20.048 & 0.009 & 20.156 & 0.011 &  \\ 
1 & 462.12 & 66.77 & 5:02:24.136 & -66:06:45.95 & 23.742 & 0.114 & 22.717 & 0.064 &  \\ 
1 & 69.32 & 67.12 & 5:02:24.129 & -66:06:28.19 & 24.354 & 0.197 & 23.408 & 0.119 &  \\ 
1 & 625.93 & 67.98 & 5:02:24.121 & -66:06:53.36 & 23.734 & 0.104 & 22.867 & 0.068 &  \\ 
1 & 481.95 & 68.05 & 5:02:24.126 & -66:06:46.85 & 26.170 & 0.933 & 24.238 & 0.229 &  \\ 
1 & 123.21 & 69.78 & 5:02:24.112 & -66:06:30.61 & 19.646 & 0.061 & 19.683 & 0.046 &  \\ 
1 & 602.06 & 70.09 & 5:02:24.107 & -66:06:52.28 & 21.049 & 0.016 & 20.913 & 0.017 &  \\ 
1 & 538.54 & 70.64 & 5:02:24.105 & -66:06:49.41 & 19.814 & 0.008 & 19.914 & 0.01 &  \\ \hline
\end{tabular}
\end{minipage}
\end{table*}

Isochrones from
Bertelli et al.\ \shortcite{Bert94}, with a range of age and
metallicity values encompassing those found in the literature for
these clusters, are plotted on the CMDs.  To illustrate the effects of
age and metallicity the top plot in Figure~\ref{fig:VIcolmag} shows 25
\& 40 Myr isochrones for two metallicities, [Fe/H], of -0.4 and 0 and
the bottom plot shows solar metallicity isochrones for ages of 25, 40
and 63 Myr.

The two clusters have very similar CMDs, which are traced well by
the 25Myr solar metallicity isochrone. The ages and metallicities
of these clusters are investigated further by comparison with
simulations in section~\ref{sec:sim}. Note that, even with
the very short exposure times used here, the brightest stars are
still saturated in the F814W images.

The red giant branch of the field population of the LMC is apparent in
the CMDs at V-I$\approx$1, V$\approx$18.5.  We have not subtracted
these background stars from our data as we are predominantly
interested in the brighter stars (V$\le$19) where the contribution
from the field is negligible (see e.\ g.\ Hunter et al.\ 1997
Figure4).

\begin{figure*}
\psfig{file=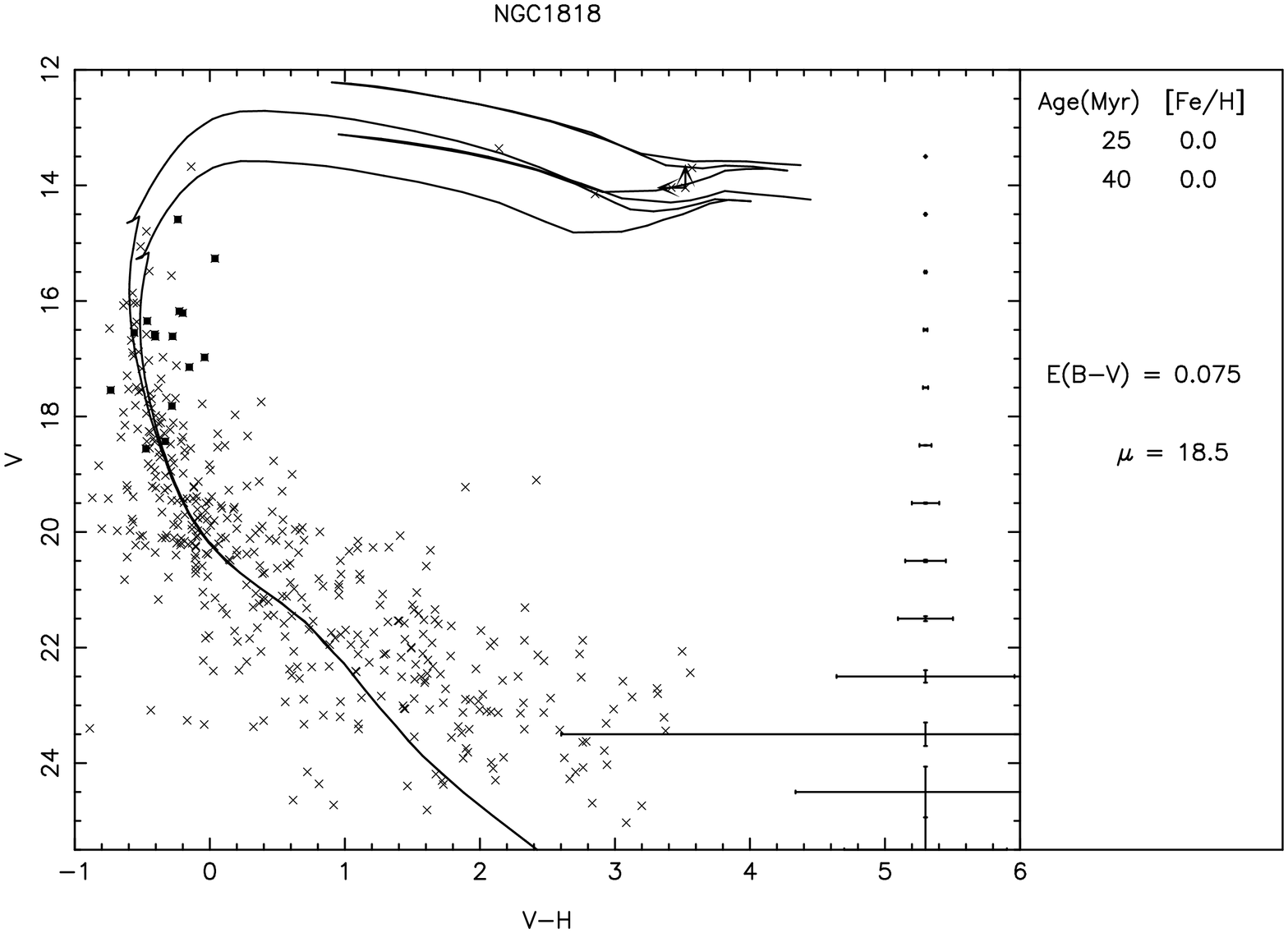,angle=-90,width=14cm}
\psfig{file=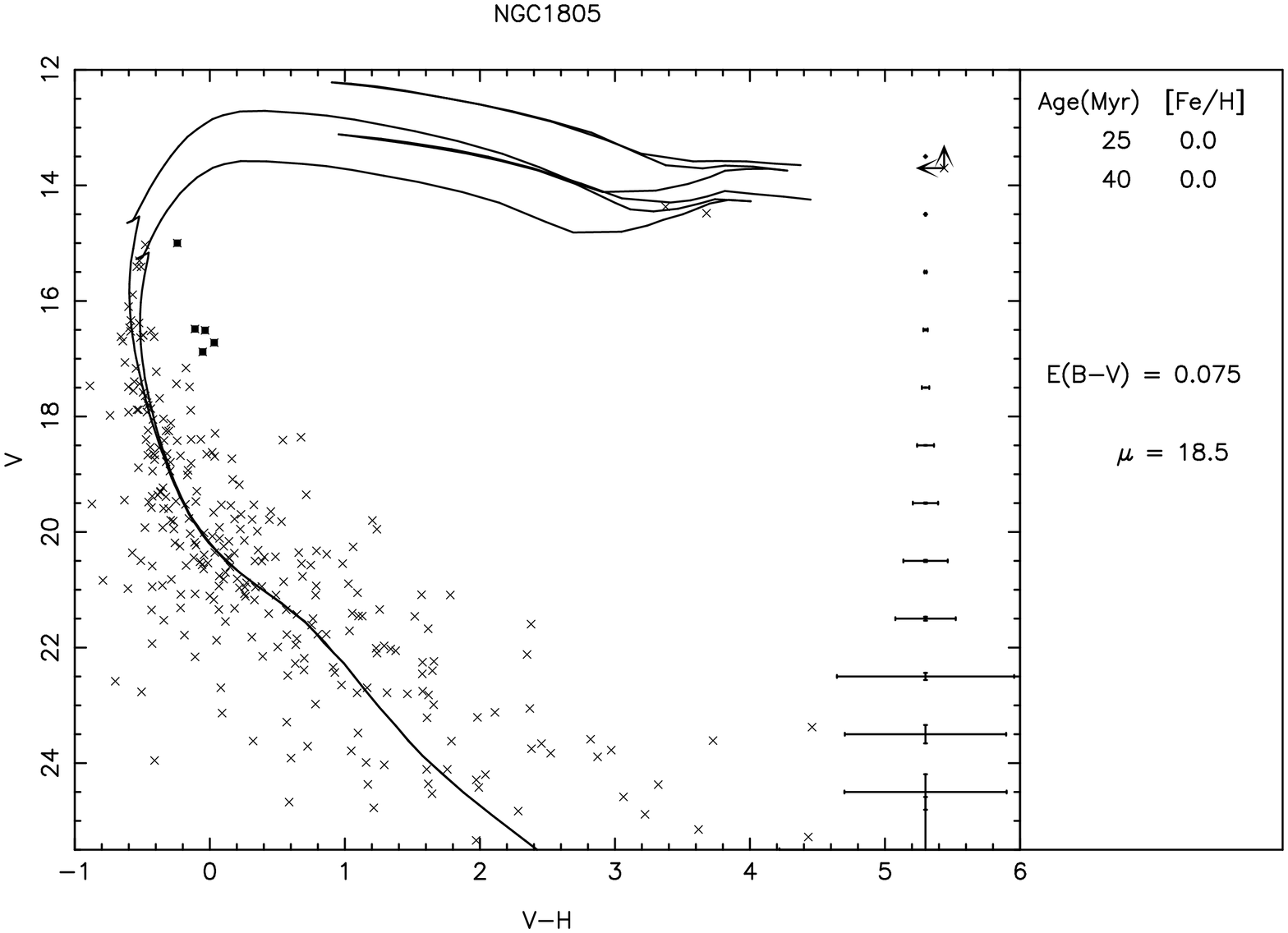,angle=-90,width=14cm}
\caption{V vs V-H colour magnitude diagrams for NGC1818 (top) and 
NGC1805 (bottom)}
\label{fig:VHcolmag}
\end{figure*}

Figure~\ref{fig:VHcolmag} show the V vs V-H diagrams for both
clusters. The isochrones \cite{Bert94} are for 25 and 40 Myr and
for solar metallicity.
Tables~\ref{tab:n1818vhdat} and \ref{tab:n1805vhdat} tabulate these data
for NGC1818 and NGC1805 respectively (full versions of these tables
are available on the MNRAS web site).
\begin{table*}
\caption{NGC1818 WFPC2 and NICMOS data table (a full version is available from the MNRAS web site). The columns contain: 1) \& 2) 
x\&y coordinates measured on the F555W image 3) \& 4) RA and Dec 5)-8) Johnson-Cousins V \& H magnitudes and errors obtained
as described in subsections~\ref{subsec:WFPC2red} \& \ref{subsec:nicred} and de-reddened using E(B-V)=0.075 9) Be star flag (see subsection~\ref{subsec:Beid})}
\label{tab:n1818vhdat}
\begin{minipage}{18cm}
\begin{tabular}{ccccccccc} \hline 
x & y &\multicolumn{2}{c}{RA (J2000) Dec} & V & $\Delta$V & H & $\Delta$H & Flag\\ \hline 
340.98 & 228.75 & 5:04:15.370 & -66:26:09.23 & 21.555 & 0.036 & 20.230 & 0.21 &  \\ 
426.83 & 229.79 & 5:04:14.905 & -66:26:11.97 & 18.832 & 0.005 & 18.830 & 0.051 &  \\ 
562.99 & 232.56 & 5:04:14.162 & -66:26:16.26 & 20.450 & 0.013 & 20.500 & 0.072 &  \\ 
484.83 & 234.05 & 5:04:14.572 & -66:26:13.70 & 999.000 & 0.045 & 21.950 & 0.219 &  \\ 
580.68 & 235.37 & 5:04:14.052 & -66:26:16.74 & 22.105 & 0.042 & 20.800 & 0.119 &  \\ 
313.66 & 236.84 & 5:04:15.473 & -66:26:08.10 & 22.506 & 0.097 & 21.080 & 0.605 &  \\ 
312.23 & 239.80 & 5:04:15.465 & -66:26:07.95 & 23.414 & 0.234 & 21.080 & 0.605 &  \\ 
511.93 & 241.08 & 5:04:14.389 & -66:26:14.35 & 20.240 & 0.011 & 20.340 & 0.06 &  \\ \hline
\end{tabular}
\end{minipage}
\end{table*}
\begin{table*}
\caption{NGC1805 WFPC2 and NICMOS data table. Columns as in Table~\ref{tab:n1818vhdat}}
\label{tab:n1805vhdat}
\begin{minipage}{18cm}
\begin{tabular}{ccccccccc} \hline 
x & y &\multicolumn{2}{c}{RA (J2000) Dec} & V & $\Delta$V & H & $\Delta$H & Flag\\ \hline 
418.22 & 213.60 & 5:02:23.046 & -66:06:43.95 & 21.550 & 0.027 & 21.434 & 0.092 &  \\ 
426.93 & 214.81 & 5:02:23.036 & -66:06:44.35 & 23.134 & 0.098 & 23.044 & 0.582 &  \\ 
404.11 & 221.83 & 5:02:22.984 & -66:06:43.31 & 17.886 & 0.003 & 18.414 & 0.009 &  \\ 
369.79 & 241.77 & 5:02:22.836 & -66:06:41.75 & 24.675 & 0.39 & 24.088 & 1.713 &  \\ 
378.92 & 242.71 & 5:02:22.829 & -66:06:42.16 & 23.486 & 0.136 & 24.543 & 2.869 &  \\ 
384.43 & 252.34 & 5:02:22.757 & -66:06:42.41 & 21.349 & 0.025 & 21.781 & 0.282 &  \\ 
368.23 & 255.18 & 5:02:22.736 & -66:06:41.68 & 21.453 & 0.025 & 20.351 & 0.091 &  \\ 
355.04 & 256.25 & 5:02:22.728 & -66:06:41.08 & 20.395 & 0.012 & 20.438 & 0.114 &  \\ \hline
\end{tabular}
\end{minipage}
\end{table*}

NGC1818 and NGC1805 have five and three red supergiants respectively.
Although our having to use NIC2 reduced the number of stars in the
NICMOS data, these colour-magnitude diagrams are still of some use as
they show us that the red supergiants lie towards the red end of the
isochrones. Although the poisson errors on the magnitudes of the
bright stars are small, there could well be systematic errors of a few
tenths of a magnitude due to uncertainty in the H calibration and the
isochrones.  In NGC1818 two of the red supergiants are located on the
25 Myr isochrone, and the other three are consistent with either the
25 or 40 Myr isochrone.  In NGC1805 two of the red supergiants are
located on the 40 Myr isochrone, and one which has lower limits in V
is consistent with either isochrone.

\subsection{Be stars}
\label{subsec:Beid}
In Figure~\ref{fig:VIcolmag} there are many stars in both clusters
with 15$\le$V$\le$17 that are significantly redder than the
isochrones. It was suspected that these are Be stars.

An effective way to identify Be stars in clusters is to use the fact
that these stars show Balmer emission and hence will separate from
non-Be stars in V-Ha `colour'.  An image of the cluster in H$\alpha$
can be used to find those stars which are H$\alpha$ bright
e.g. \cite{Greb92}.  An archive H$\alpha$ image exists for NGC1818
\cite{Kell00} and this has been used to identify the Be stars. The
H$\alpha$ image is not registered with the V image and so not all the
stars have H$\alpha$ data.  The single pointing H$\alpha$ image
contains many cosmic rays so it is not possible to detect stars
independently on this image.  The coordinate transform between our
data and the H$\alpha$ image was found and the star coordinates were
transformed to the H$\alpha$ image and aperture magnitudes obtained.
Each star position on the H$\alpha$ image was checked to see if the
H$\alpha$ magnitude was contaminated by a cosmic ray. Cosmic rays were
identified by their brightness and morphology.  For a few stars the
H$\alpha$ image did contain a cosmic ray at the star position and it
was not possible to obtain an H$\alpha$ magnitude.

\begin{figure}
\psfig{file=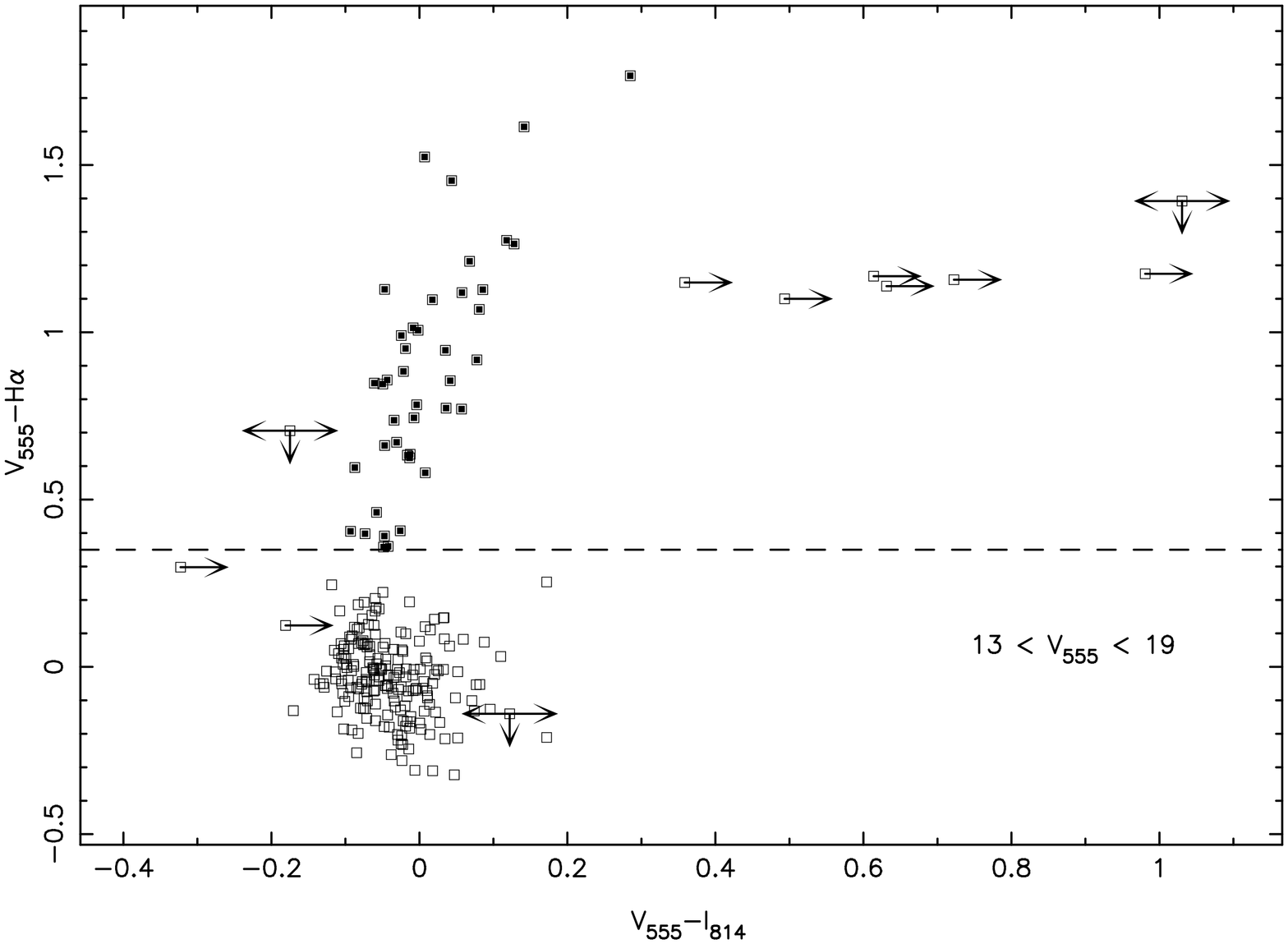,angle=-90,width=9cm}
\caption{Example of Be star identification for stars on the PC chip in the V image. Be stars are the filled points with V$_{555}$-H$\alpha>$0.35. The points with V$_{555}$-I$_{814}>$0.3 are red supergiants.}
\label{fig:beid}
\end{figure}

Figure~\ref{fig:beid} shows an example of the plots used to identify the
Be stars. 
Any star with V$_{555}$-H$\alpha$ greater than that of the bulk of the
stars, that does not lie in the red supergiant region of the
colour-magnitude diagram, is identified as a Be star.  Red supergiant
stars are excluded as they can also show H$\alpha$ emission. We have
looked at stars with 13$<$V$_{555}\!<$19 and a star is identified as a
Be star if V$_{555}$-H$\alpha\!>$0.35.  The red supergiant region of the
CMD is defined to be V$_{555}$-I$_{814}\!>$0.3 for V$_{555}\!<$16 and
V$_{555}$-I$_{814}\!>$0.6 for 16$<$V$_{555}\!<$19.

Unfortunately, no archive H$\alpha$ image exists for NGC1805 and so we
cannot identify the Be stars in this cluster using the method
above. However, almost all the stars in NGC1818 with V$_{555}\!<$17.5
and 0$<$V$_{555}$-I$_{814}\!<$0.5 are Be stars. Since the colour-magnitude
diagrams for the two clusters appear very similar we have assumed that
the same is true in NGC1805. NGC1818 does contain Be stars that are
outside the above limits in V and V-I that are intermingled with
non-Be stars. It is likely therefore that our colour-magnitude
diagrams of NGC1805 contain some unidentified Be stars. 

Table~\ref{tab:bestars} shows the number of Be stars and all stars
found in 0.5 magnitude bins for each cluster. The first two columns
give the numbers found using the methods discussed above (H$\alpha$
image for NGC1818 and region in the CMD for NGC1805). The final column
gives the number of Be stars in NGC1818 using the same Be star
criteria as in NGC1805.
\begin{table}
\caption{Numbers of Be and all (Be+non-Be stars) in 0.5 magnitude bins. The Be star numbers in the NGC1818 H$\alpha$ column are from identification of Be stars using an H$\alpha$ narrowband image. In NGC1805 stars are classified as Be stars if they are in a region of the colour-magnitude diagram which only contains Be stars in NGC1818. The final NGC1818 column shows the number of Be stars in NGC1818 using the same classification criteria as in NGC1805. See text for more details.}
\begin{tabular}{lllllll} \hline
V & \multicolumn{2}{l}{NGC1818 H$\alpha$} & \multicolumn{2}{l}{NGC1805} & \multicolumn{2}{l}{NGC1818} \\ 
      & N(Be) & N(all) & N(Be) & N(all) & N(Be) & N(all) \\ \hline
14.25 & 1 & 2  & 2 & 2  & 1 & 2 \\
14.75 & 2 & 4  & 2 & 3  & 1 & 5 \\
15.25 & 1 & 8  & 0 & 9  & 1 & 11\\
15.75 & 7 & 21 & 0 & 8  & 7 & 29\\
16.25 & 9 & 29 & 1 & 14 & 5 & 36\\
16.75 & 7 & 40 & 5 & 31 & 4 & 61\\
17.25 & 4 & 34 & 2 & 36 & 5 & 60\\
17.75 & 4 & 54 & - & -  & - & - \\
18.25 & 8 & 81 & - & -  & - & - \\
18.75 & 2 & 107& - & -  & - & - \\ \hline
\end{tabular}
\label{tab:bestars}
\end{table}

\begin{figure}
\psfig{file=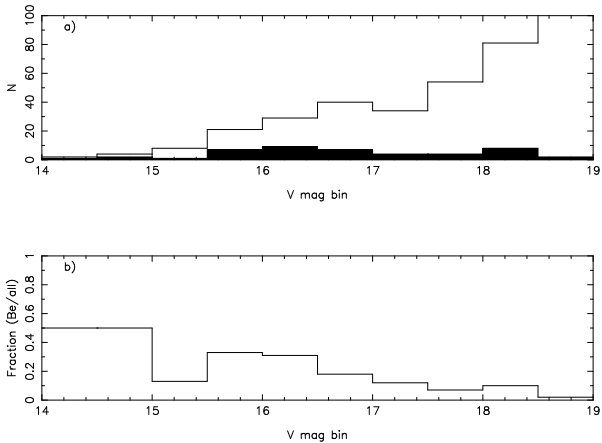,width=6.4cm,angle=-90}
\caption{a) Histograms of the number of Be stars (filled) and of all (Be+non-Be) stars in 0.5 magnitude bins down the main sequence of NGC1818. b) the ratio of Be stars to all stars in 0.5 magnitude bins down the main-sequence of NGC1818}
\label{fig:NGC1818befrac}
\end{figure}

Figure~\ref{fig:NGC1818befrac} compares the number of Be stars to the
total number of main sequence stars (including the Be stars) for the
centre of NGC1818 on the PC chip. In this and following figures the
V magnitudes are de-reddened Johnson-Cousins magnitudes.
As also found in Keller et al.\ \shortcite{Kell00}, the Be star fraction
peaks at the brightest magnitudes around the turn-off and then falls
off.

\begin{figure}
\psfig{file=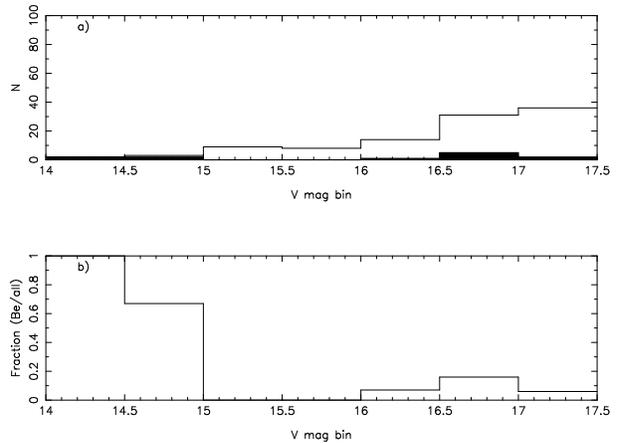,width=6.4cm,angle=-90}
\caption{As Figure~\ref{fig:NGC1818befrac} but for NGC1805. Note that in this cluster the number of Be stars in each 0.5 magnitude bin, and hence the Be star fraction, is a lower limit (see text).}
\label{fig:NGC1805befrac}
\end{figure}

\begin{figure}
\psfig{file=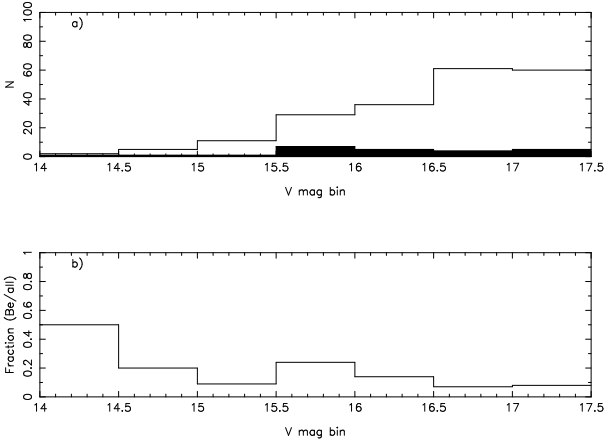,width=6.4cm,angle=-90}
\caption{As Figure~\ref{fig:NGC1818befrac}, except that here we
analyse NGC1818, using the same criteria for Be star definition as
used for NGC1805 in Figure~\ref{fig:NGC1805befrac}. Note that the
total number of stars in the bins is slightly higher in this figure
than in Figure~\ref{fig:NGC1818befrac}. This is because the H$\alpha$
image used for Be star identification in NGC1818 did not align exactly
with the PC chip.}
\label{fig:NGC1818befrac1}
\end{figure}

Figure~\ref{fig:NGC1805befrac} show the Be star fraction for the centre of
NGC1805 on the PC chip. 
Because the Be stars in NGC1805 were identified using the criteria
V$_{555}<$17.5 and V$_{555}$-I$_{814}>$0 the numbers of Be stars and
Be star fractions shown in Figure~\ref{fig:NGC1805befrac} are lower
limits. Again there is a peak in the Be star fraction around the
turn-off.  The most significant difference in the Be star fractions of
NGC1818 and NGC1805 is the lack of Be stars in the range 15$<$V$<$16
in the latter.  To see whether this is just due to the different
selection criteria used for identifying Be stars inNGC1805 we have
used the same criteria to select Be stars in NGC1818.  The results are
shown in Figure~\ref{fig:NGC1818befrac1}.

The fraction of Be stars with 15$<$V$<$16 in NGC1818 using these
selection criteria is $\approx$0.2. This fraction of Be stars in
NGC1805 would give $\approx$ 3 Be stars in NGC1805, whereas we observe
none.  There is therefore tentative evidence for a difference in Be
star fraction between NGC1818 and NGC1805, but this needs to be
investigated further by obtaining an H$\alpha$ HST image for NGC1805
to allow proper Be star identification.

\begin{figure}
\psfig{file=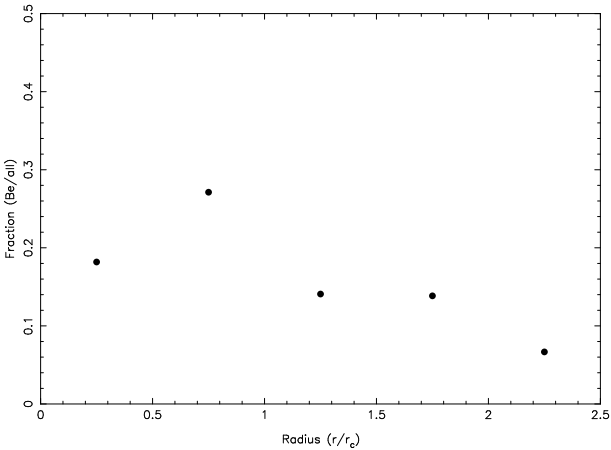,width=6.5cm,angle=-90}
\caption{Change of Be star fraction with radius in NGC1818. Stars with 
14$<$V$<$19 are considered.}
\label{fig:NGC1818beradfrac}
\end{figure}

Figure~\ref{fig:NGC1818beradfrac} shows the change in Be star fraction with
radius in NGC1818. This is for stars on the PC with 14$<$V$<$19.
It can be seen that the Be star fraction decreases with distance from
the cluster centre. This is consistent with the mass segregation of
the bright stars (V$<\approx$18) in NGC1818 found by Elson et al.\
\shortcite{Elso98}.

\section{Simulations}
\label{sec:sim}
There are several populations contributing to the appearance of the
bright end of the colour magnitude diagram in these young clusters -
main sequence stars, Be stars, binaries, evolved supergiant stars and
blue stragglers.  These various contributors complicate finding the
age, metallicity and reddening for the clusters. For instance blue
stragglers can make a cluster appear younger than it really is e.g.
\cite{Greb96} and binaries produce a spread at
the top of the main sequence that looks similar to an age spread.  In
order to take these various affects into account we have compared the
observations with simulations of clusters of various ages and
metallicities. In this section we first discuss the simulations and
then compare them with the observations to investigate the cluster
parameters.

Synthetic colour-magnitude diagrams are generated using the rapid
evolution code developed by Hurley, Pols \& Tout \shortcite{Hurl00a}
which covers all aspects of the evolution from the main-sequence up
to, and including, the remnant stages.  Binary evolution is accounted
for by incorporating the algorithm described by Hurley
\shortcite{Hurl00}.  This model, which supercedes the work of Tout et
al.\ \shortcite{Tout97}, includes tidal circularization and
synchronization, angular momentum loss mechanisms, mass transfer,
common-envelope evolution, collisions and supernova kicks.  These
evolution algorithms allow realistic CMDs to be developed accurately
and efficiently, for any age and for all metallicities in the range
$10^{-4}$ to 0.03.  At present these simulations do not allow for
stellar rotation, which is known to move massive stars redward in the
colour magnitude diagram \cite{Meyn00}.  The synthetic CMDs are
particularly useful for comparison with observed clusters whose
stellar populations have not been significantly altered by dynamical
interactions, such as the young clusters described here.  In the case
of dense, or dynamically old, clusters the interaction between cluster
environment and the evolution of the constituent stars must be
consistently taken into account when simulating CMDs.  For this reason
the stellar and binary evolution algorithms have been incorporated
into a state-of-the-art $N$-body code \cite{Aars99,Hurl00b} but this
is not used in this work.

To compare with the data we have run simulations with ages of 10, 25
and 40 Myr and metallicities Z = 0.01 and 0.02. These ages
and metallicities encompass the literature values and are also
implied by the isochrones shown in Figure~\ref{fig:VIcolmag}. There are
approximately 850000 stars per simulation with masses down to 0.1
M$_{\odot}$.  The initial mass function for single stars is taken from
Kroupa, Tout \& Gilmore \shortcite{Krou93}.  The simulated binary
fraction is 35\%, with binary masses taken from the initial mass
function of Kroupa, Tout \& Gilmore \shortcite{Krou91} as this has
not been corrected for the effects of binaries, and a uniform
distribution of mass-ratios.

\begin{figure*}
\psfig{file=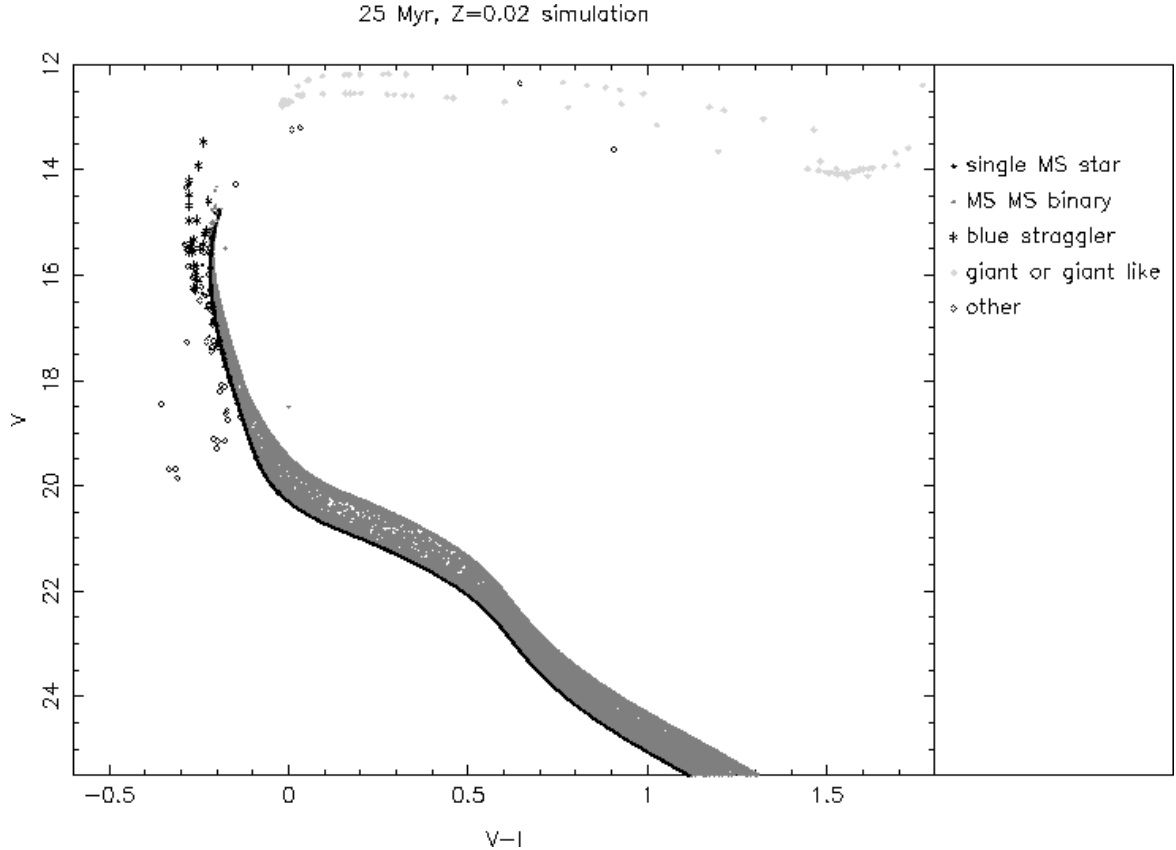,width=11.2cm,angle=-90}
\caption{A simulated cluster colour-magnitude diagram for
an age of 25 Myr and a metallicity Z=0.02. This plot contains
$\approx$425000 simulated stars.}
\label{fig:25sim}
\end{figure*}

Figure~\ref{fig:25sim} plots the simulated CMD for an age of 25Myr
metallicity Z=0.02. Here we just show half ($\approx$425000) the total
number of simulated stars.
The blue stragglers are plotted with asterisks. They are defined in the
simulation as main-sequence stars with a mass $> 1.02 \times$turn-off
mass. This definition does not identify blue stragglers in binaries.
Figure~\ref{fig:25sim} shows that, as well as sitting above the
turn-off in the region where blue stragglers have been found
observationally, there are also simulated blue stragglers below and
blueward of the turn-off.  These less luminous blue stragglers are the
result of binary interactions of stars below the turn-off.  If present
in sufficient numbers in a cluster they could cause a widening of the
main sequence around the turn-off in the colour-magnitude diagrams.
This would then complicate the search for age spreads amongst the
massive stars as these also produce a spread in colour at a given
magnitude.

The absolute numbers of blue stragglers in the simulations are somewhat
arbitrary, as the blue straggler numbers are most likely affected by
the cluster structure and dynamical evolution in ways that are
currently not well understood. We therefore use the simulations to
find the position of blue stragglers in the colour-magnitude diagrams
and hence locate possible blue stragglers in the real clusters.

To add the observed errors to the simulations we have modelled the
observed error distribution, and found the model that makes the colour
spread in the observed and simulated CMDs the same for 18$\le$V$<$20.

From each simulation we find the expected number of stars in each
cluster.  The simulation is normalised to the data using the number of
observed and simulated stars with 17$<$V$<$19.  This magnitude range
avoids background contamination and also the bright star region where
the blue straggler fraction is somewhat uncertain.

\begin{figure*}
\psfig{file=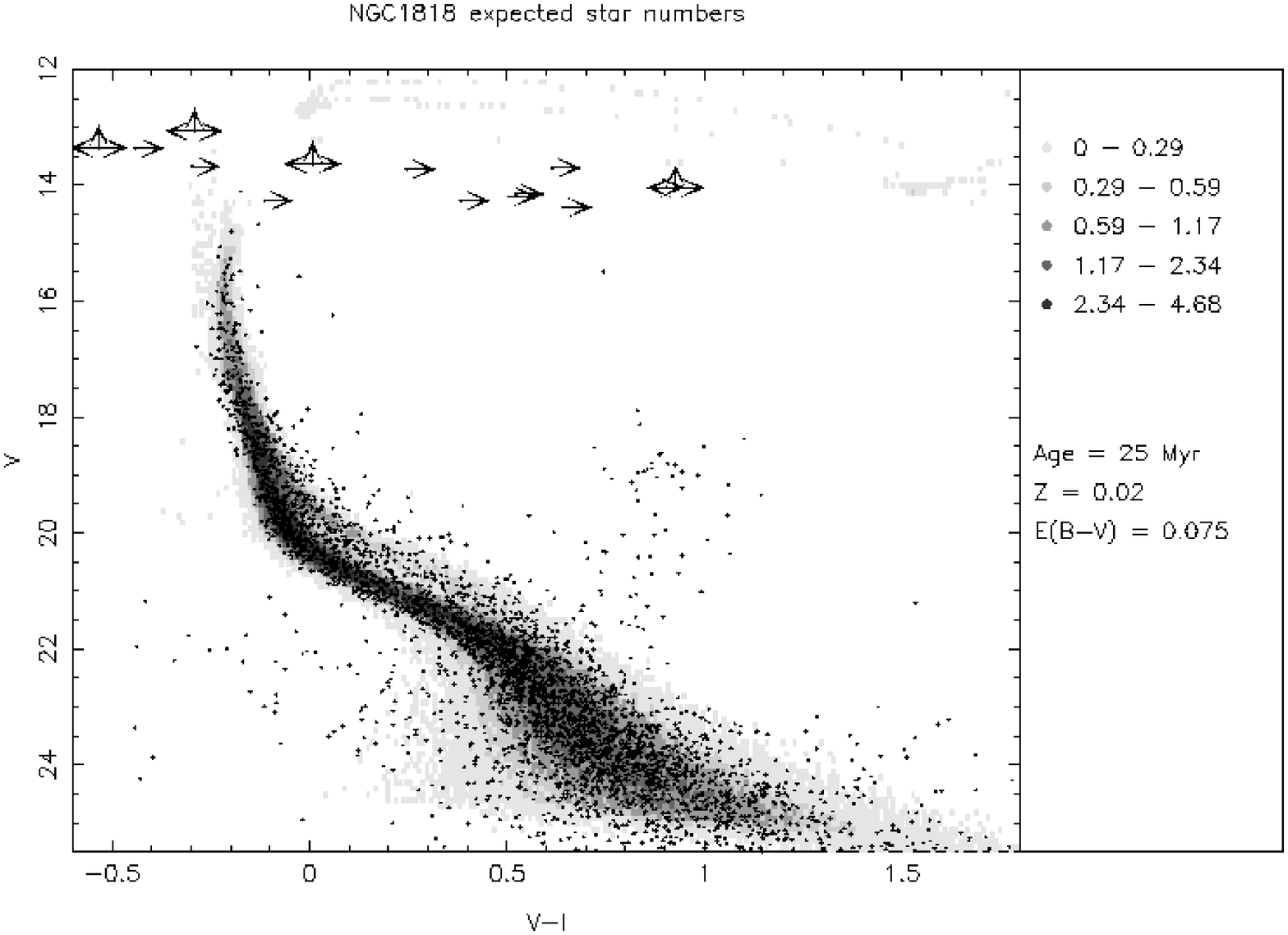,width=11.2cm,angle=-90}
\psfig{file=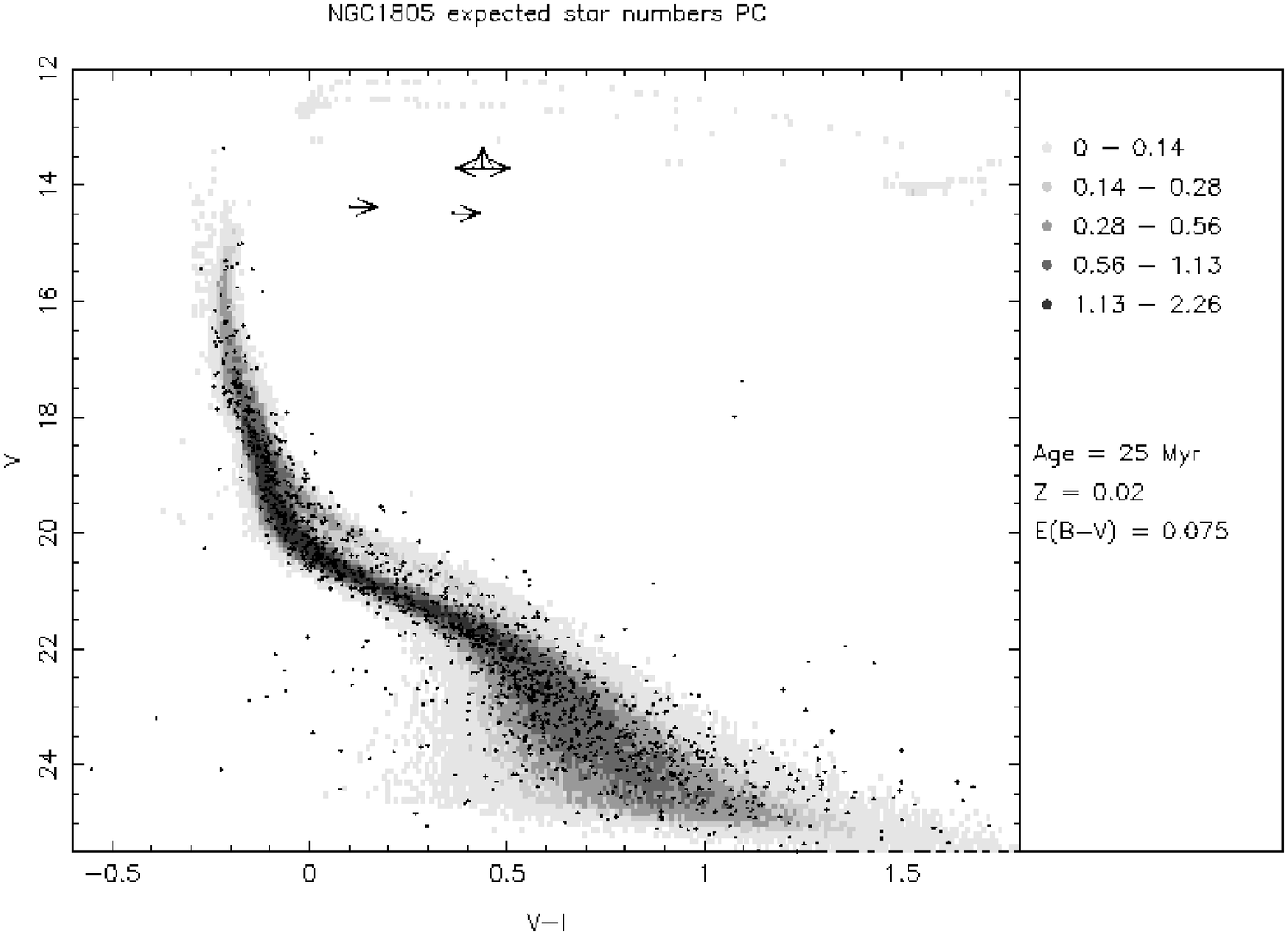,width=11.2cm,angle=-90}
\caption{Comparison of simulated (greyscale) and real data for NGC1805 (top) and NGC1818 (bottom). The simulations have a metallicity Z=0.02 and an
age of 25 Myr. Both clusters are fit well by this simulation for 
V$>$19 where the shape depends only on the metallicity.}
\label{fig:metsim}
\end{figure*}

First we find the metallicity that best describes the cluster CMDs.
As can be seen from the isochrones in Figure~\ref{fig:VIcolmag}, for
V$>$19 the isochrone shape depends only on metallicity and not on age.
Also, the reddening vector is virtually parallel to the main sequence
in this region and so reddening also does not significantly affect the
fit.  We have checked whether the background LMC field population
could be influencing our metallicity determination using the
colour-magnitude diagram in Figures 3 \& 8 of Hunter et al.\
\shortcite{Hunt97}. These show that there is very little difference
between the pre and post background subtraction colour-magnitude
diagrams at all magnitudes, and for stars brighter than V$\approx$21
there is no difference. Figure~\ref{fig:metsim} shows the observed and
simulated CMDs of NGC1818 and NGC1805. In NGC1805 there is a small
colour shift between the chips (see Section~\ref{sec:data}), and so we
just compare the simulations to the colour-magnitude diagram from the
PC chip, which contains most of the bright cluster stars. The points
are the observed data and the grey scale is the simulation. The grey
scale gives the expected number of stars in a box of width (V-I) 0.01
mags and height (V) 0.1 mags.  The simulation has age 25 Myr,
metallicity Z=0.02 and the data have been de-reddened assuming
E(B-V)=0.075.  For ease of comparison the Be stars have been removed
from the real data in Figure~\ref{fig:metsim}.Both of the clusters are
described well by the simulation for V$>$19.

The solar metallicity that we find here is higher than the
[Fe/H]$\approx$-0.4 that is expected for the LMC clusters (see
discussion in section~\ref{sec:intro}).  It could be the case that
these clusters are more metal-rich than the surrounding population as
there are no metallicity measurements for stars in NGC1805 and only a
couple of stars measured with high resolution in NGC1818.  Some of the
difference could however be due to a systematic error in the
calibration of HST magnitudes to Johnson-Cousins which, as noted in
subsection~\ref{subsec:WFPC2red} could be as high as 0.04 in V-I.

Next we fix the simulation metallicity at Z=0.02 and consider the age
of each cluster.  The age affects the appearance of the CMD at bright
mags (V$<$18).  We have compared each cluster to 10, 25 and 40 Myr old
simulations, and also to combined simulations of two ages, 10\&25 Myr
and 25\&40 Myr. The latter are used to provide constraints on any age
spread in the clusters.

The isochrones in Figure~\ref{fig:VIcolmag} illustrate that the red
supergiant positions in the CMD are mostly affected by age.  The
observed red supergiant numbers and magnitudes rule out the 10 Myr old
simulation.  Unfortunately, as can also be seen from
Figure~\ref{fig:VIcolmag}, the exposure time for the F814W
observations was long enough to saturate these stars and so the red
supergiant positions do not allow us to distinguish between the 25 and
40 Myr old simulations.  Most of the red supergiants are not saturated
in the V vs V-H colour-magnitude diagrams in
Figure~\ref{fig:VHcolmag}. In these CMDs the red supergiant positions
are consistent with ages of 25 Myr or 25\&40 Myr for NGC1818 and 40
Myr or 25\&40 Myr for NGC1805.

To compare the observations and the different age simulations we first
look at the shape of the observed and expected CMDs.  The reddening
also affects the position of the CMD at bright magnitudes, and so we
have compared simulations and observations for 25, 40, 10\&25 and
25\&40 Myr and E(B-V) between 0 and 1.  We find that only the 25 Myr
and 25\&40 Myr simulations with E(B-V)=0.075, describe well the
observed shape. To distinguish between these two simulations we look
at the observed and expected numbers of bright stars in 0.5 mag bins
down the main sequence.

Table~\ref{tab:brightnum} gives the observed and expected numbers of
the brightest stars for the 25 and 25\&40 Myr simulations for
12$<$V$<$17. Recall that the simulations and observations were forced
to have the same number of stars in the range 17$<$V$<$19 used for
normalisation. The observed numbers here include the Be stars.  The
simulations do not contain Be stars, but the Be phenomenon just moves
stars redwards in the colour-magnitude diagram, and so the numbers per
magnitude bin remain the same.
\begin{table}
\caption{Comparison of the bright star observed and simulated numbers
in 0.5 mag bins. The columns are as follows: 1) V mag of bin centre 2)
observed number 3) \& 4) expected number in 25 and 25\&40 simulation
5) \& 6) number of observed stars that fall in blue straggler region
in each simulation.  The numbers in brackets in 1) are stars with V
lower limits. Simulation numbers in grey boxes do not fit the observed
numbers, those in clear boxes only fit the observed numbers if some of
the observed stars are blue stragglers (i.e. the number of observed
blue stragglers is greater than in the simulation.)}
\begin{tabular}{llllll} \hline
Mag bin & Obs.\ No. &\multicolumn{2}{c}{Sim predicted no.}&\multicolumn{2}{c}{No. possible BS} \\ \hline
      &       &25 &25\&40&25&25\&40 \\
\multicolumn{6}{l}{NGC1818} \\
12.25 & 0     & 0.89 & 0.50 & 0 & 0 \\
12.75 & 0     & 2.44 & 1.24 & 0 & 0 \\
13.25 & 3 (2) & 0.28 & 2.24 & 0 & 0 \\
13.75 & 4 (1) & 0.87 & 0.76 & 1 & 1 \\
14.25 & 9 (1) & \fcolorbox{black}{yellow}{1.73} & \fcolorbox{black}{yellow}{1.50} & 0 & 0 \\
14.75 & 4     & 2.47 & 3.07 & 0 & 0 \\
15.25 & 8     & 7.25 & 5.38 & 2 & 3 \\
15.75 & 21    &\fcolorbox{black}{yellow}{13.28} &\fbox{11.71} & 1 & 8 \\ 
16.25 & 29    &\fcolorbox{black}{yellow}{18.37} &\fbox{18.22} & 3 &12 \\
16.75 & 40    &\fcolorbox{black}{yellow}{27.66} &\fcolorbox{black}{yellow}{27.25} & 0 & 2 \\
\multicolumn{6}{l}{NGC1805} \\
12.25 & 0     & 0.43 & 0.24 & 0 & 0 \\
12.75 & 0     & 1.18 & 0.60 & 0 & 0 \\
13.25 & 1     & 0.13 & 1.08 & 1 & 0 \\
13.75 & 1 (1) & 0.42 & 0.36 & 0 & 0 \\
14.25 & 4     & \fcolorbox{black}{yellow}{0.83} & \fcolorbox{black}{yellow}{0.72} & 0 & 0 \\
14.75 & 3     & 1.19 & 1.48 & 0 & 0 \\
15.25 & 8     & \fcolorbox{black}{yellow}{3.50} & \fbox{2.59} & 1 & 4 \\
15.75 & 3     & \fcolorbox{black}{yellow}{6.40} & \fcolorbox{black}{yellow}{5.65} & 1 & 1 \\
16.25 & 8     & 8.86 & 8.78 & 1 & 6 \\
16.75 &18     & \fcolorbox{black}{yellow}{13.34}&\fbox{13.14} & 0 & 2 \\ \hline
\end{tabular}
\label{tab:brightnum}
\end{table}

The comparison of the observed and simulated numbers is complicated by
the fact that the number of blue stragglers in the simulations is
somewhat arbitrary. From the simulations we expect $\approx$1 blue
straggler per cluster, but there could be more than this.  The
simulations provide the location of blue stragglers in the CMDs. For
each magnitude bin we have noted how many observed stars fall in the
simulation region that contains blue stragglers i.e.  the maximum
number of observed stars that could be blue stragglers in that
magnitude bin.  Columns 5 and 6 in Table~\ref{tab:brightnum} give this
number.

The fit of the simulation to the observation is deemed to be
acceptable if the simulated number is within 1$\sigma$ of the observed
number (where $\sigma$ is the poisson error on the observed number,
which is much bigger than the poisson error on the simulated number).
The simulated numbers in the grey boxes are those that do not fit to
the observed number. The simulated numbers in the clear boxes only fit
if some of the observed stars are blue stragglers, and are therefore
not included in the simulated number.

Mass segregation is also seen in NGC1818 \cite{Elso98} but this will
not have a significant effect on the observed number of stars, as the
colour-magnitude diagram is made from all the chips and so includes
the inner and outer region of the cluster.  In NGC1805, where we are
just comparing the PC with the simulations we do not see any evidence
for mass segregation.

From Table~\ref{tab:brightnum} we find that in both clusters the
25\&40 Myr simulation is a better fit to the observed numbers than the
25 Myr simulation. In both clusters some of the observed stars must be
blue stragglers in order for the 25\&40 Myr simulation to fit.  Also in
both clusters there are a couple of magnitude bins where the
simulation does not fit even if the maximum possible number of blue
stragglers are present in the observations.

In section~\ref{sec:sim} we noted that the objects defined as blue
stragglers in our simulations are a superset of those that are
observationally called blue stragglers.  Observational blue straggers
are those that are brighter and bluer than the main sequence turn-off.
Neither NGC1818 nor NGC1805 contains such an observational blue
straggler sequence.  The simulations also contained less luminous blue
stragglers that lie just below and blueward of the turn-off.  It is
these less luminous blue stragglers that allow us to obtain a better
fit if they are present in the cluster.

The grey-scale CMD simulations in Figure~\ref{fig:metsim} show the
well-known blue Hertzsprung gap. This gap at 13$<$V$<$14 and
V-I$\approx$-0.3 is due to a very fast evolutionary stage. Stars pass
through this region of the colour-magnitude diagram comparatively
quickly and hence very few are expected to be observed there. However,
Keller et al.\ \shortcite{Kell00} find stars located in this region in
the LMC clusters they study (NGC330, NGC1818, NGC2004 and NGC2100).
We find three possible blue Hertzsprung gap stars (all with I lower
limits) in NGC1818 (consistent with Keller et al. who find two) and
one star in this region in NGC1805.  Keller et al.\ have effective
temperatures for their blue Hertzsprung gap stars and therefore they
rule out the possibility that the stars are blue stragglers, and
suggest instead that they may be due to internal mixing in
main-sequence stars in excess of that predicted by standard overshoot
models.

\section{Discussion and Conclusions}
\label{sec:conc}

We have analysed HST V,I (WFPC2) and H (NICMOS2) observations of two
young LMC clusters NGC1818 and NGC1805.  The colour-magnitude diagrams
of both clusters appear very similar.

We identify Be stars in NGC1818 using an archive H$\alpha$ image and
find that the fraction of Be stars decreases with radius from the
cluster centre. This is consistent with the mass segregation known to
exist in this cluster. There is some evidence that the fraction of Be
stars in NGC1805 may differ from that in NGC1818, but this requires
further investigation.

NGC1818 and NGC1805 do not contain large blue straggler populations.
NGC1818 contains three stars that are located in the blue Hertzsprung
gap (two of these are just consistent with the location of blue
stragglers) and NGC1805 contains one star in this region. The star in
NGC1805 also lies on the edge of the blue straggler region. 
However these blue Hertzsprung gap stars are most likely not blue
stragglers \cite{Kell00}. Further observations of blue stragglers in
clusters of all ages, in parallel with N-body simulations, are required
to understand the parameters which influence the blue straggler
fraction.

To obtain the cluster parameters (age and metallicity) we have
compared the cluster colour-magnitude diagrams to sophisticated
stellar population simulations. Comparing with simulations of a single
age we find that both clusters are best fit with solar metallicity and
an age of 25 Myr.  The fit is improved if the simulations contain
$\approx$equal masses of stars of 25 \& 40 Myr. This possible age
spread is $\approx$ a few dynamical times in these clusters.  Such a
timescale does not allow us to constrain the efficiency of the star
formation process as both high and low efficiency will leave a bound
cluster (assuming that the LMC clusters are young analogs to the
Galactic globulars). High efficiency is suggested by the similarity of
observed density profiles in LMC clusters \cite{Elso87}.

Our simulations show that, as well as lying above and bluewards of the
turnoff, blue stragglers also appear just below and blueward of the
turn-off, where they do not stand out observationally, but do increase
the width and number density of stars in those regions of the
colour-magnitude diagram.  The exact numbers of blue stragglers in the
simulations is somewhat arbitrary and we find that the best fit is
obtained between the simulations and observations if there are
significant numbers of these fainter blue stragglers in both clusters.

\section*{acknowledgements}
We would like to thank Stefan Keller for telling us about the archival
H$\alpha$ images.

\end{document}